\documentclass[11pt]{article}
 \usepackage{latexsym,epic,eepic}
\usepackage{amssymb}
\usepackage{amsmath}
\usepackage{amsfonts}
\usepackage{graphicx}

\begin{document}
\newcommand{\2}{\vspace{0.2 cm}}
\newcommand{\dist}{{\rm dist}}
\newcommand{\diam}{{\rm diam}}
\newcommand{\rad}{{\rm rad}}
\newcommand{\dom}{\mbox{$\rightarrow$}}
\newcommand{\ndom}{\mbox{$\not\rightarrow$}}
\newcommand{\sdom}{\mbox{$\Rightarrow$}}
\newcommand{\nsdom}{\mbox{$\not\Rightarrow$}}
\newcommand{\qed}{\hfill$\diamond$}
\newcommand{\pf}{{\bf Proof: }}
\newtheorem{theorem}{Theorem}[section]
\newcommand{\ra}{\rangle}
\newcommand{\la}{\langle}
\newtheorem{lemma}[theorem]{Lemma}
\newtheorem{claim}[theorem]{Claim}
\newtheorem{definition}[theorem]{Definition}
\newtheorem{corollary}[theorem]{Corollary}
\newtheorem{proposition}[theorem]{Proposition}
\newtheorem{conjecture}[theorem]{Conjecture}
\newtheorem{problem}[theorem]{Problem}
\newtheorem{remark}[theorem]{Remark}
\newtheorem{example}[theorem]{Example}
\newcommand{\beq}{\begin{equation}}
\newcommand{\eeq}{\end{equation}}
\newcommand{\argmax}{{\rm argmax}}
\newcommand{\MiP}{MinHOM($H$) }
\newcommand{\MaP}{MaxHOM($H$) }
\newcommand{\vecc}[1]{\stackrel{\leftrightarrow}{#1}}

\title{Minimum Cost Homomorphisms \\ to Proper Interval Graphs and Bigraphs}

\date{}

\author{Gregory Gutin\thanks{Corresponding author. Department of Computer Science,
Royal Holloway University of London, Egham, Surrey TW20 OEX, UK,
gutin@cs.rhul.ac.uk and Department of Computer Science, University
of Haifa, Israel}  \and Pavol Hell\thanks{School of Computing
Science, Simon Fraser University, Burnaby, B.C., Canada, V5A 1S6,
pavol@cs.sfu.ca} \and Arash Rafiey\thanks{Department of Computer
Science, Royal Holloway University of London, Egham, Surrey TW20
OEX, UK, arash@cs.rhul.ac.uk} \and Anders Yeo\thanks{Department of
Computer Science, Royal Holloway University of London, Egham,
Surrey TW20 OEX, UK, anders@cs.rhul.ac.uk}} \maketitle

\begin{abstract}
For graphs $G$ and $H$, a mapping $f:\ V(G)\dom V(H)$ is a
homomorphism of $G$ to $H$ if $uv\in E(G)$ implies $f(u)f(v)\in
E(H).$ If, moreover, each vertex $u \in V(G)$ is associated with
costs $c_i(u), i \in V(H)$, then the cost of the homomorphism $f$
is $\sum_{u\in V(G)}c_{f(u)}(u)$. For each fixed graph $H$, we
have the {\em minimum cost homomorphism problem}, written as
MinHOM($H)$. The problem is to decide, for an input graph $G$ with
costs $c_i(u),$ $u \in V(G), i\in V(H)$, whether there exists a
homomorphism of $G$ to $H$ and, if one exists, to find one of
minimum cost. Minimum cost homomorphism problems encompass (or are
related to) many well studied optimization problems. We describe a
dichotomy of the minimum cost homomorphism problems for graphs
$H$, with loops allowed. When each connected component of $H$ is
either a reflexive proper interval graph or an irreflexive proper
interval bigraph, the problem MinHOM($H)$ is polynomial time
solvable. In all other cases the problem MinHOM($H)$ is NP-hard.
This solves an open problem from an earlier paper. Along the way,
we prove a new characterization of the class of proper interval
bigraphs.
\end{abstract}

\section{Motivation and Terminology}

We consider finite undirected and directed graphs without multiple
edges, but with loops allowed. For a directed or undirected graph
$H$, $V(H)$ ($E(H)$) denotes the set of vertices (edges) of $G.$
We will reserve the term 'graph' for undirected graphs and use the
term 'digraph' for directed graphs. A graph or digraph without
loops will be called {\em irreflexive}; a graph or digraph in
which every vertex has a loop will be called {\em reflexive}.

The {\em intersection graph} of a family ${\cal
F}=\{S_1,S_2,\ldots,S_n\}$ of sets is the graph $G$ with
$V(G)={\cal F}$ in which $S_i$ and $S_j$ are adjacent just if
$S_i\cap S_j\neq \emptyset$. Note that by this definition, each
intersection graph is reflexive. (This is not the usual
interpretation \cite{golumbic2004,spin}.) A graph isomorphic to
the intersection graph of a family of intervals on the real line
is called an {\em interval graph}. If the intervals can be chosen
to be inclusion-free, the graph is called a {\em proper interval
graph}. Thus both interval graphs and proper interval graphs are
reflexive. The {\em intersection bigraph} of two families ${\cal
F}_1=\{S_1,S_2,\ldots,S_n\}$ and  ${\cal
F}_2=\{T_1,T_2,\ldots,T_m\}$ of sets is the bipartite graph with
$V(G)={\cal F}_1 \cup {\cal F}_2$ in which  $S_i$ and $T_j$ are
adjacent just if $S_i\cap T_j\neq \emptyset$. Note that by this
definition an intersection bigraph is irreflexive (as are all
bipartite graphs). A bipartite graph isomorphic to the
intersection bigraph of two families of intervals on the real line
is called an {\em interval bigraph}. If the intervals in each
family ${\cal F}_i$ can be chosen to be inclusion-free, the graph
is called a {\em proper interval bigraph}. Thus both interval
bigraphs and proper interval bigraphs are irreflexive.

For directed or undirected graphs $G$ and $H$, a mapping $f:\
V(G)\dom V(H)$ is a {\em homomorphism of $G$ to $H$} if $uv\in
E(G)$ implies $f(u)f(v)\in E(H).$ Recent treatments of
homomorphisms in directed and undirected graphs can be found in
\cite{hell2003,hell2004}. Let $H$ be a fixed directed or
undirected graph. The {\em homomorphism problem} for $H$ asks
whether a directed or undirected input graph $G$ admits a
homomorphism to $H.$ The {\em list homomorphism problem} for $H$
asks whether a directed or undirected input graph $G$ with lists
(sets) $L_u \subseteq V(H), u \in V(G)$ admits a homomorphism $f$
to $H$ in which all $f(u) \in L_u, u \in V(G)$.

There have been several studies of homomorphism (and more
generally constraint satisfaction) problems with costs. Most
frequently, it is only the edges $ij$ of the graph $H$ that have
costs, and $H$ is then taken to be a complete (reflexive) graph
\cite{anzhu,alon}. In this context, one seeks a homomorphism $f$
of the input graph $G$ to $H$ for which the sum, over all $uv \in
E(G)$, of the costs of $f(u)f(v)$ is minimized. These are typified
by problems such as finding a maximum bipartite subgraph, or, in
the context of more general constraints, finding an assignment
satisfying a maximum number of clauses \cite{alon}. More
generally, \cite{cohenJAIR22} considers instead of costs of edges
$ij$ of $H$, the costs of mapping an edge $uv$ of $G$ to an edge
$ij$ of $H$. In this way, the constraint on the edge $uv$ is
`soft' - it may map to any pair $ij$ of $H$, but with cost that
depends both on $uv$ and on $ij$. Nonbinary constraints are
treated the same way in \cite{cohenJAIR22}. This general `soft'
constraint satisfaction context of \cite{cohenJAIR22} allows for
vertex weights as well, since they can be viewed as unary
constraints. Nevertheless, in combinatorial optimization it makes
sense to investigate vertex weights alone, insisting that binary
(and higher order) constraints are hard (or `crisp'). This is the
path we take, focusing on problems in which each possible
assignment of a value to a variable has an associated cost.

We now formulate our problem, in the context of graph
homomorphisms. (Of course, there is a natural counterpart for
constraint satisfaction problems in general.) Suppose $G$ and $H$
are directed (or undirected) graphs, and $c_i(u)$, $u\in V(G)$,
$i\in V(H)$ are nonnegative {\em costs}. The {\em cost of a
homomorphism} $f$ of $G$ to $H$ is $\sum_{u\in V(G)}c_{f(u)}(u)$.
If $H$ is fixed, the {\em minimum cost homomorphism problem},
MinHOM($H$), for $H$ is the following decision problem. Given an
input graph $G$, together with costs $c_i(u)$, $u\in V(G)$, $i\in
V(H)$, and an integer $k$, decide if $G$ admits a homomorphism to
$H$ of cost not exceeding $k$. Informally, we also use MinHOM($H$)
to denote the corresponding optimization problem in which we want
to minimize the cost of a homomorphism of $G$ to $H$, or state
that none exists. The minimum cost of a homomorphism of $G$ to $H$
(if one exists) will be denoted by mch$(G,H)$. For simplicity, we
shall always assume the graph $G$ to be irreflexive. (Note that if
we have to solve a problem in which some vertices $u$ have loops,
we can account for the loops by changing the weights $c_i(u)$ to
be infinite on all vertices $i$ of $H$ which do not have a loop.)

The problem MinHOM$(H)$ was introduced in \cite{gutinDAMlora},
where it was motivated by a real-world problem in defence
logistics. We believe it offers a practical and natural model for
optimization of weighted homomorphisms. It is easy to see that the
homomorphism problem (for $H$) is a special case of MinHOM$(H)$,
obtained by setting all weights to $0$ (and taking $k=0$).
Similarly, the list homomorphism problem (for $H$) is obtained by
setting $c_i(u)=0$ if $i \in L_u$ and $c_i(u)=1$ otherwise (and
taking $k=0$). When $H$ is an irreflexive complete graph, the
problem MinHOM$(H)$ becomes the so-called {\em general optimum
cost chromatic partition} problem, which has been intensively
studied \cite{halld2001,jansenJA34,jiangGT32}, and has a number of
applications, \cite{kroon1997,supowitCAD6}. Two special cases of
that problem that have been singled out are the {\em optimum cost
chromatic partition problem}, obtained when all $c_i(u), u \in
V(G)$, are the same (the cost only depends on the colour $i$)
\cite{kroon1997}, and the {\em chromatic sum problem}, obtained
when each $c_i(u)=i$ (the cost of the colour $i$ is the value $i$,
i.e., we are trying to minimize the sum of the assigned colours)
\cite{jiangGT32}.

For the homomorphism problem for undirected graphs $H$ (with loops
allowed), the following dichotomy classification is known: if $H$
is bipartite or has a loop, the problem is polynomial time
solvable; otherwise it is NP-complete \cite{hellJCT48}. For the
list homomorphism problem for undirected graphs $H$ (with loops
allowed), a similar dichotomy classification is also known
\cite{biarc}. None of the weighted versions of homomorphism
problems cited above has a known dichotomy classification. This
includes the soft constraint satisfaction problem of
\cite{cohenJAIR22}, even though the authors identify a class of
polynomially solvable constraints that is in a certain sense
maximal. We shall provide a dichotomy classification of the
complexity of MinHOM$(H)$.

Preliminary results on MinHOM$(H)$ for irreflexive graphs were
obtained by Gutin, Rafiey, Yeo and Tso in \cite{gutinDAMlora}: it
was shown there that MinHOM($H$) is polynomial time solvable if
$H$ is an irreflexive bipartite graph whose complement is an
interval graph, and NP-complete when $H$ is either a nonbipartite
graph or a bipartite graph whose complement is not a circular arc
graph. This left as unclassified a large class of irreflexive
graphs, settled in this paper. In fact, we shall provide a general
classification which applies to graphs with loops allowed.

\begin{theorem}\label{maint}
Let $H$ be a connected graph (with loops allowed). If $H$ is a
proper interval graph or a proper interval bigraph, then the
problem MinHOM($H$) is polynomial time solvable. In all other
cases, the problem MinHOM($H$) is NP-complete.
\end{theorem}

We note that in the two polynomial cases, the graph $H$ is either
irreflexive or reflexive. Indeed, it is easy to see that if $H$
contains an edge $rs$ where $r$ has a loop and $s$ doesn't, then
the problem MinHOM($H$) is NP-complete. It suffices to notice that
if $G$ has all vertex costs $c_s(u)=0, u \in V(G)$, and all other
vertex costs $c_i(u)=1, u \in V(G), i \neq s$, then there exists a
homomorphism of cost not exceeding $k$ if and only if $G$ has an
independent set of size $|V(G)|-k$. Thus it suffices to consider
the reflexive and irreflexive cases separately, as we shall do in
the remainder of the paper.

\begin{corollary}
Let $H$ be any graph (with loops allowed). If each component of
$H$ is a (reflexive) proper interval graph or an (irreflexive)
proper interval bigraph, then the problem MinHOM($H$) is
polynomial time solvable. In all other cases, the problem
MinHOM($H$) is NP-complete.
\end{corollary}

\pf Indeed, if $H$ is not connected, it consists of components
$H_i$, and we may solve, for every component $G_j$ of $G$, all
problems MinHOM$(H_i)$ in turn, and then take the smallest cost
homomorphism amongst these, as the solution for the component
$G_j$. (Note that a homomorphism of $G$ is described by giving a
homomorphism for every $G_j$.) \qed

\vspace{2mm}

%A more general problem of minimum cost {\em soft} homomorphisms
%was introduced in \cite{cohenJAIR22}. The problem was stated more
%generally for constraint satisfaction problems, but focusing on graphs,
%we may reformulate it as follows. Suppose $H$ is a fixed {\em set}. Given
%an input graph $G$, together with nonnegative costs $c_i(u), u \in V(G)$,
%$i \in V(H)$, and nonnegative costs $c_{ij}(uv), uv \in E(G)$, $i, j \in V(H)$,
%find a mapping $f$ of $V(G)$ to $H$ which minimizes the sum $\sum_{u\in
%V(G)}c_{f(u)}(u) + \sum_{uv \in E(G)} c_{f(u)f(v)}(uv)$. Note that our
%problem MinHOM$(H)$ for a {\em graph} $H$ is obtained by setting the
%`soft' constraints to the `hard' (or `crisp') values $c_{f(u)f(v)}(uv)=0$
%if $f(u)f(v) \in E(H)$ and $c_{f(u)f(v)}(uv)=\infty$ if $f(u)f(v) \not\in E(H)$.

\vspace{2mm} Finally, in Section \ref{frsec}, we discuss the
situation for digraphs. It turns out that there are digraphs that
do not have the Min-Max property but which have polynomially
solvable problems MinHOM$(H)$. At this point the classification is
open, although we do mention some partial results.

\section{Polynomial Algorithms} \label{polysec}

%When the binary costs $c_{ij}(uv)$ are {\em submodular functions}, the
%problem of minimum cost soft homomorphisms was solved in \cite{cohenJAIR22}.
%For our problem MinHOM$(H)$, where $H$ is a digraph, the submodularity
%condition was simplified in \cite{gutinDAM} as follows.

We say that a digraph $H$ has the {\em Min-Max property} if its
vertices can be ordered $w_1,w_2, \ldots, w_p$ so that if $i < j$,
$s < r$ and $w_iw_r, w_jw_s \in E(H)$, then $w_iw_s \in E(H)$ and
$w_jw_r \in E(H)$.

This property was first defined in \cite{gutinDAM}, where is was
identified as an important property of digraphs, as far as the
problem MinHOM$(H)$ is concerned. (We should point out that the
original definition, which is easily seen equivalent to the one
given above, required that if $w_iw_r, w_jw_s \in E(H)$, then also
$w_xw_y \in E(H)$ for $x=\min(i,j), y=\min(r,s)$ and for
$x=\max(i,j),$ $y=\max(r,s)$.)

Using an algorithm of \cite{cohenJAIR22}, the authors of
\cite{gutinDAM} proved the following result. (The proof in
\cite{gutinDAM} is only stated for irreflexive digraphs, but it is
literally the same for digraphs in general.)

\begin{theorem}\cite{gutinDAM}\label{MMd}
Let $H$ be a digraph. If $H$ satisfies the Min-Max property, then
MinHOM($H$) is polynomial time solvable.\end{theorem}

The Min-Max property is very closely related to a property of
digraphs that has been of interest since \cite{welzl}. We say that
a digraph $G$ has the {\em $X$-underbar property} if its vertices
can be ordered $w_1,w_2,$ $ \ldots, w_p$ so that if $i < j$, $s <
r$ and $w_iw_r, w_jw_s \in E(H)$, then $w_iw_s \in E(H)$. (In
other words, $w_iw_r, w_jw_s \in E(H)$ implies that $w_xw_y \in
E(H)$ for $x=\min(i,j), y=\min(r,s)$). It is interesting to note
that the $X$-underbar property already ensures that the list
homomorphism problem for $H$ has a polynomial solution
\cite{welzl,hell2004}.

We first apply Theorem \ref{MMd} to reflexive graphs. It is
important to keep in mind that we may view undirected graphs as
digraphs, by replacing each edge $uv$ of the undirected graph by
the two opposite edges $uv, vu$ of the directed graph; this does
not affect which mappings are homomorphisms \cite{hell2004}. Under
this interpretation, we observe the following.

\begin{proposition} \label{refl}
A reflexive graph $H$ has the Min-Max property if and only if its
vertices can be ordered $w_1,w_2, \ldots, w_p$ so that $i < j < k$
and $w_iw_k \in E(H)$ imply that $w_iw_j \in E(H)$ and $w_jw_k \in
E(H)$.
\end{proposition}

\pf To see that the condition is necessary, consider the directed
edge $w_iw_k$ and the loop $w_jw_j$ and apply the definition in
digraphs. To see that it is sufficient, suppose $i < j$, $s < r$
and $w_iw_r, w_jw_s \in E(H)$. Observe that, up to symmetry, there
are only two nontrivial cases possible - typified by $s < i < r <
j$ and $s < i < j < r$. In both cases, the condition in the
theorem and the loops $w_iw_i$ and $w_rw_r$ (respectively
$w_jw_j$) ensure that $w_iw_s \in E(H)$ and $w_jw_r \in E(H)$.
\qed

\vspace{2mm}

The condition in Proposition \ref{refl} is known to characterize
proper interval graphs \cite{certi,spin}.

\begin{corollary} \label{umbre}
A reflexive graph $H$ has the Min-Max property if and only if it
is a proper interval graph.
\end{corollary}

For irreflexive graphs $H$, we observe that the standard view of
$H$ as a digraph will not work. Indeed, if both $uv$ and $vu$ are
directed edges of the digraph $H$, then the Min-Max property
requires that both $uu$ and $vv$ be loops of $H$. Therefore, we
shall view a bipartite graph $H$, with a fixed bipartition into
(say) {\em white} and {\em black} vertices, as a digraph in which
all edges are directed from white to black vertices. Under this
interpretation, we observe the following fact. (We have simply
replaced one ordering of all vertices with the induced orderings
on white and black vertices; note that given orderings of white
and black vertices, any total ordering preserving the relative
orders of white and of black vertices satisfies the condition.)

\begin{proposition} \label{irrefl}
A bipartite digraph $H$, with a fixed bipartition into white and
black vertices, and with all edges oriented from white to black
vertices, has the Min-Max property if and only if the white
vertices can be ordered as $u_1,u_2,\ldots,u_p$ and the black
vertices can be ordered as $v_1,v_2,\ldots,v_q$, so that if $i <
j$, $s < r$ and $u_iv_r, u_jv_s \in E(H)$, then $u_iv_s \in E(H)$
and $u_jv_r \in E(H)$. \qed
\end{proposition}

We now remark that this condition is in fact a previously unknown
equivalent definition of proper interval bigraphs. (There are many
such equivalent definitions, see \cite{hellJGT,spin}.)

\begin{proposition} \label{kotz}
A bipartite graph $H$, with a fixed bipartition into white and
black vertices, is a proper interval bigraph if and only if the
white vertices can be ordered as $u_1,u_2,\ldots,u_p$ and the
black vertices can be ordered as $v_1,v_2,\ldots,v_q$, so that if
$i < j$, $s < r$ and $u_iv_r, u_jv_s \in E(H)$, then $u_iv_s \in
E(H)$ and $u_jv_r \in E(H)$.
\end{proposition}

\pf Suppose $H$ is isomorphic to the intersection bigraph of the
family ${\cal F}_1$ of white intervals and the family ${\cal F}_2$
of black intervals, each being inclusion-free. We can order the
white and the black vertices of $H$ (corresponding to the white
and black intervals respectively), by the left to right order of
the intervals. (Since the intervals in each family are
inclusion-free, this order is uniquely defined by either the left
or right endpoints of the intervals.) We now claim that these
orders $u_1,u_2,\ldots,u_p$ and $v_1,v_2,\ldots,v_q$ satisfy the
above property. Thus suppose that $i < j$ and $s < r$, and
$u_iv_r, u_jv_s \in E(H)$. This means that the white interval
$U_i$ corresponding to $u_i$ intersects the black interval $V_r$
corresponding to $v_r$, and the white interval $U_j$ corresponding
to $u_j$ intersects the black interval $V_s$ corresponding to
$v_s$. Since the interval $U_j$ to the right of $U_i$ and the
interval $V_s$ to the left of $V_r$, this means that $U_i$ must
also intersect $V_s$ and $U_j$ must also intersect $V_r$.

Conversely, suppose that we have the white and black vertices of
$H$ ordered as $u_1,u_2,\ldots,u_p$ and $v_1,v_2,\ldots,v_q$, so
that the claimed property holds. Define, for each white vertex
$u_i$, its {\em leftmost} and {\em rightmost} neighbours $L(i),
R(i)$ respectively, as the smallest respectively largest subscript
$x$ with $u_iv_x \in E(H)$. It follows from the stated property
that if $i < j$ then $L(i) \leq L(j)$ and $R(i) \leq R(j)$.
Moreover, $u_iv_k\in E(H)$ if and only if $L(i)\le k \le R(i)$.
Indeed, suppose that $L(i)\le k\le R(i)$, but $u_iv_k\not\in
E(H)$. We may assume that $v_k$ is not an isolated vertex and
$u_jv_k$ with $j>i$. The stated property implies that $u_iv_k\in
E(H)$, a contradiction. We now define two families of intervals
$U_i, i=1,2,\dots,p$ and $V_j, j=1,2,\dots,q$ as follows. Each
$V_j$ will be the interval $[j-\frac{1}{4},j+\frac{1}{4}]$. Each
$U_i$ will be the interval
$[L(i)-\frac{1}{2^i},R(i)+\frac{1}{2}-\frac{1}{2^i}]$. It is easy
to see that $U_i, V_j$ intersect if and only if $u_iv_j \in E(H)$.
Because of the above properties of $L(i), R(i)$, and because of
the small fractions perturbing the intervals, the two families are
inclusion-free. \qed

\vspace{2mm}

It now follows that we can apply Theorem \ref{MMd} to reflexive
proper interval graphs and irreflexive bipartite proper interval
bigraphs, to deduce the polynomial algorithms in Theorem
\ref{maint}.

\begin{corollary}
If $H$ is a proper interval graph or a proper interval bigraph,
then the problem MinHOM($H$) is polynomial time solvable.
\end{corollary}

\pf For proper interval graphs $H$ this directly follows from
Theorem \ref{MMd}, Proposition \ref{refl}, and Corollary
\ref{umbre}. For proper interval bigraphs, we shall note that we
may assume that the graph $G$ is also bipartite, else no
homomorphism to $H$ exists. We may also assume that $G$ is
connected, as otherwise we can solve the problem for each
component separately. Thus we may take $G$ to be given with white
and black vertices (only two such partitions are possible for a
connected graph), and orient all edges from white to black
vertices. Now we can use Theorem \ref{MMd}, and Propositions
\ref{irrefl} and \ref{kotz}, to derive a polynomial solution. \qed

\vspace{2mm}

We shall now describe the polynomial time algorithms. They follow
from \cite{cohenJAIR22}, via the translation in \cite{gutinDAM},
which depends on submodularity of the cost functions, allowing the
problem to be decomposed into a series of minimum weight cut
problems. We now show how, in our case, one can solve the problem
directly as a single minimum weight cut problem. (This is similar
to what is done in \cite{cohenJAIR22} for a related situation.)
For simplicity, we shall focus on the reflexive case, although the
technique applies to irreflexive graphs as well.

Thus suppose that $H$ is a reflexive proper interval graph, with
vertices ordered $w_1, w_2,\dots, w_p$, so that $i<j<k$ and
$w_iw_k \in E(H)$ imply $w_iw_j \in E(H)$ and $w_jw_k \in E(H)$.
For simplicity we shall write $i$ instead of $w_i$. We denote, for
each $i$, by $\ell(i)$ the largest subscript $j$ such that $j < i$
and $j$ is not adjacent to $i$, if such a $j$ exists. Note for
future reference that if $i' \leq i$, then $i'$ is adjacent to $i$
if and only if $\ell(i) < i'$.

Given a graph $G$ with costs $c_i(u), u \in V(G), i \in V(H)$, we
construct an auxiliary digraph $G \times H$ as follows. The vertex
set of $G \times H$ is $V(G) \times V(H)$ together with two other
vertices, denoted by $s$ and $t$. The directed weighted edges of
$G \times H$ are

\begin{itemize}
\item an edge from $s$ to $(u,1)$, of weight $\infty$, for each $u
\in V(G)$,

\item an edge from $(u,i)$ to $(u,i+1)$, of weight $c_i(u)$, for
each $u \in V(G)$ and $i \in V(H)$,

\item an edge from $(u,p)$ to $t$, of weight $c_p(u)$, for each $u
\in V(G)$, and

\item an edge from $(u,i)$ to $(v,\ell(i))$, of weight $\infty$,
for every edge $uv \in E(G)$ and each $i \in V(H)$ such that
$\ell(i)$ is defined.
\end{itemize}

\noindent (Note that each undirected edge $uv$ of $G$ gives rise
to two directed edges $(u,i)(v,\ell(i))$ and $(v,i)(u,\ell(i))$,
both of infinite weight, in the last statement.)

\vspace{2mm} A {\em cut} in $G \times H$ is a partition of the
vertices into two sets $S$ and $T$ such that $s \in S$ and $t \in
T$; the weight of a cut is the sum of weights of all edges going
from a vertex of $S$ to a vertex of $T$. Let $S$ be a cut of
minimum (finite) weight, and define $j_u$ to be the maximum value
such that $(u,j_u) \in S$. Let $S'$ be the cut containing $s$ and
all $(u,1), (u,2), \dots, (u,j_u)$, for all $u \in V(G)$. If $S'
\not= S$, then either the weight of $S'$ is infinite, or at most
that of $S$, as the only arcs we might add to the cut are of the
form $(u,i)(v,l(i))$. If the weight of $S'$ is infinite, then
there must be an arc of the form $(u,i)(v,\ell(i))$ in the cut
$S'$, where neither $(u,i)$ nor $(v,\ell(i))$ belong to $S$. Note
that $\ell(i)>j_v$ as $(v,\ell(i)) \not\in S'$. Furthermore
$\ell(j_u) \geq \ell(i)$, as $j_u>i$, which implies that
$\ell(j_u)>j_v$. Therefore the edge $(u,j_u)(v,\ell(j_u))$
belonged to the cut $S$, which therefore had infinite weight, a
contradiction. Therefore $S'=S$. Now define a  mapping $f$ from
$V(G)$ to $V(H)$ by setting $f(u)=j_u$. This must be a
homomorphism of $G$ to $H$; indeed, suppose that $uv\in E(G)$, but
$j_uj_v\not\in E(H)$. Without loss of generality assume that $j_v
\leq j_u$, which implies that $j_v \leq \ell(j_u)$. This implies
that the edge $(u,j_u)(v,\ell(j_u))$ belongs to the cut $S$, a
contradiction. Conversely, any minimum cost homomorphism $f$ of
$G$ to $H$ corresponds, in this way, to a minimum weight cut of $G
\times H$.

We conclude that the minimum weight of cut in $G \times H$ is
exactly equal to the minimum cost of a homomorphism of $G$ to $H$.
Since minimum weighted cuts can be found by standard flow
techniques, we obtain a polynomial time algorithm. Specifically,
we note that the graph $G \times H$ has $O(|V(G)||V(H)|)$
vertices. Using the best minimum cut (maximum flow) algorithms, we
obtain minimum cost homomorphisms in time $O(|V(G)|^3|V(H)|^3)$
\cite{schr}; if $H$ is fixed, and $G$ has $n$ vertices, this is
$O(n^3)$.

We observe that this sort of product construction is also similar
to the algorithm in \cite{gutinDAM}, which transforms the minimum
cost homomorphism problem into a maximum independent set problem
in another kind of product $G \otimes H$. Note that these kinds of
algorithms, which solve the problem via a product construction
involving $G$ and $H$, are polynomial even if $H$ is part of the
input.

\section{NP-completeness Proofs}\label{NPsec}

In this section it will be more convenient to begin with the
irreflexive case. Hence all graphs are irreflexive unless stated
otherwise.

A bipartite graph $H$ with vertices $x_1,x_2,x_3,x_4,y_1,y_2,y_3$
is called
\begin{description}
\item{\em a bipartite claw} if
$E(H)=\{x_4y_1,y_1x_1,x_4y_2,y_2x_2, x_4y_3, y_3x_3\};$

\item{\em a bipartite net} if
$E(H)=\{x_1y_1,y_1x_3,y_1x_4,x_3y_2,x_4y_2,y_2x_2,y_3x_4\};$

\item{\em a bipartite tent} if
$E(H)=\{x_1y_1,y_1x_3,y_1x_4,x_3y_2,x_4y_2,y_2x_2,y_3x_4\}.$
\end{description}
See Figure \ref{III_graphs}.

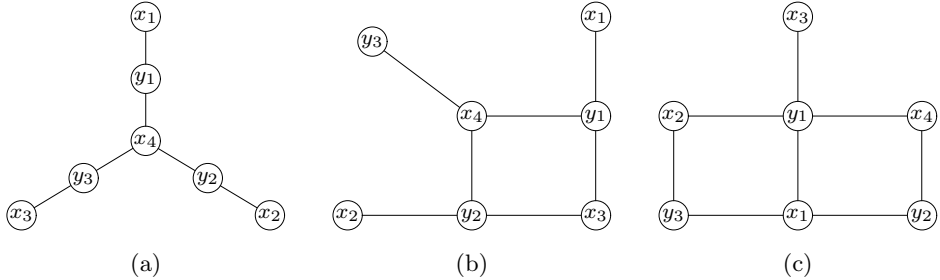
\begin{figure}
\unitlength 0.330mm \linethickness{0.4pt} \noindent
\begin{picture}(   120.00,   100.00)
\put(    60.00,    90.00){\circle{12.0}} \put(  60.000,
90.000){\makebox(0,0){{\scriptsize $x_1$}}} \put(   110.00,
10.00){\circle{12.0}} \put( 110.000,
10.000){\makebox(0,0){{\scriptsize $x_2$}}} \put(    10.00,
10.00){\circle{12.0}} \put(  10.000,
10.000){\makebox(0,0){{\scriptsize $x_3$}}} \put(    60.00,
40.00){\circle{12.0}} \put(  60.000,
40.000){\makebox(0,0){{\scriptsize $x_4$}}} \put(    60.00,
65.00){\circle{12.0}} \put(  60.000,
65.000){\makebox(0,0){{\scriptsize $y_1$}}} \put(    85.00,
25.00){\circle{12.0}} \put(  85.000,
25.000){\makebox(0,0){{\scriptsize $y_2$}}} \put(    35.00,
25.00){\circle{12.0}} \put(  35.000,
25.000){\makebox(0,0){{\scriptsize $y_3$}}} \drawline(  60.000,
84.000)(  60.000,  71.000) \drawline(  60.000,  59.000)(  60.000,
46.000) \drawline( 104.855,  13.087)(  90.145,  21.913) \drawline(
79.855,  28.087)(  65.145,  36.913) \drawline(  15.145,  13.087)(
29.855,  21.913) \drawline(  40.145,  28.087)(  54.855,  36.913)
\put(60,-10){\makebox(0,0){{\footnotesize (a)}}}
\end{picture} \hspace{-0.01cm}
\linethickness{0.4pt}
\begin{picture}(   120.00,   100.00)
\put(   110.00,    90.00){\circle{12.0}} \put( 110.000,
90.000){\makebox(0,0){{\scriptsize $x_1$}}} \put(    10.00,
10.00){\circle{12.0}} \put(  10.000,
10.000){\makebox(0,0){{\scriptsize $x_2$}}} \put(   110.00,
10.00){\circle{12.0}} \put( 110.000,
10.000){\makebox(0,0){{\scriptsize $x_3$}}} \put(    60.00,
50.00){\circle{12.0}} \put(  60.000,
50.000){\makebox(0,0){{\scriptsize $x_4$}}} \put(   110.00,
50.00){\circle{12.0}} \put( 110.000,
50.000){\makebox(0,0){{\scriptsize $y_1$}}} \put(    60.00,
10.00){\circle{12.0}} \put(  60.000,
10.000){\makebox(0,0){{\scriptsize $y_2$}}} \put(    20.00,
80.00){\circle{12.0}} \put(  20.000,
80.000){\makebox(0,0){{\scriptsize $y_3$}}} \drawline( 110.000,
84.000)( 110.000,  56.000) \drawline( 110.000,  44.000)( 110.000,
16.000) \drawline( 104.000,  10.000)(  66.000,  10.000) \drawline(
60.000,  16.000)(  60.000,  44.000) \drawline(  66.000,  50.000)(
104.000,  50.000) \drawline(  16.000,  10.000)(  54.000,  10.000)
\drawline(  24.800,  76.400)(  55.200,  53.600)
\put(60,-10){\makebox(0,0){{\footnotesize (b)}}}
\end{picture}  \hspace{-0.01cm}
\linethickness{0.4pt}
\begin{picture}(   120.00,   100.00)
\put(    60.00,    10.00){\circle{12.0}} \put(  60.000,
10.000){\makebox(0,0){{\scriptsize $x_1$}}} \put(    10.00,
50.00){\circle{12.0}} \put(  10.000,
50.000){\makebox(0,0){{\scriptsize $x_2$}}} \put(    60.00,
90.00){\circle{12.0}} \put(  60.000,
90.000){\makebox(0,0){{\scriptsize $x_3$}}} \put(   110.00,
50.00){\circle{12.0}} \put( 110.000,
50.000){\makebox(0,0){{\scriptsize $x_4$}}} \put(    60.00,
50.00){\circle{12.0}} \put(  60.000,
50.000){\makebox(0,0){{\scriptsize $y_1$}}} \put(   110.00,
10.00){\circle{12.0}} \put( 110.000,
10.000){\makebox(0,0){{\scriptsize $y_2$}}} \put(    10.00,
10.00){\circle{12.0}} \put(  10.000,
10.000){\makebox(0,0){{\scriptsize $y_3$}}} \drawline(  60.000,
16.000)(  60.000,  44.000) \drawline(  60.000,  56.000)(  60.000,
84.000) \drawline(  16.000,  50.000)(  54.000,  50.000) \drawline(
10.000,  44.000)(  10.000,  16.000) \drawline( 104.000,  50.000)(
66.000,  50.000) \drawline( 110.000,  44.000)( 110.000,  16.000)
\drawline(  54.000,  10.000)(  16.000,  10.000) \drawline( 66.000,
10.000)( 104.000,  10.000)
\put(60,-10){\makebox(0,0){{\footnotesize (c)}}}
\end{picture}

\mbox{ }

\caption{A bipartite claw  (a), a bipartite net (b) and a
bipartite tent (c).}
\end{figure}  \label{III_graphs}

These graphs play an important role for proper interval bigraphs.
One of the equivalent characterizations is the following
\cite{hellJGT}.

\begin{theorem}\label{mainnp}\cite{hellJGT} A bipartite graph $H$ is
a proper interval bigraph if and only if it does not contain an
induced cycle of length at least six, or a bipartite claw, or a
bipartite net, or a bipartite tent.
\end{theorem}

It follows that to show that MinHOM($H$) is NP-complete when $H$
is not a proper interval bigraph, it suffices to prove that
MinHOM($H$) is NP-complete when $H$ is either a cycle of length at
least six, or a bipartite claw, or a bipartite net, or a bipartite
tent. Indeed, if MinHOM$(H)$ is NP-complete and $H$ is an induced
subgraph of $H'$, then MinHOM$(H')$ is also NP-complete, as we may
set the costs $c_i(u)=\infty$ for all vertices $u$ of $G$ and all
$i$ which are vertices of $H'$ but not of $H$. The NP-completeness
of MinHOM$(H)$ for bipartite cycles of length at least six follows
from \cite{federC19}. In the remainder of this section, we prove
that MinHOM($H$) is NP-complete for the bipartite claw, net, and
tent.

\vspace{2mm}

We shall use the following tool.

\begin{theorem} \label{MaxIndepNPhard}
The problem of finding a maximum independent set in a $3$-partite
graph $G$ (even given the three partite sets) is NP-complete.
\end{theorem}

\pf
  Let ${\cal G}_3$ be the set of all graphs of degree at most 3
with at least three vertices excluding $K_4.$ By the well-known
theorem of Brooks (see, e.g., \cite{west1996}), every graph in
${\cal G}_3$ is 3-partite. Using Lovasz' constructive proof of
Brooks' theorem in \cite{lovaszJCT19}, one can find three partite
sets of a graph $G\in {\cal G}_3$ in polynomial time.

Nevertheless, Alekseev and Lozin  showed recently in
\cite{alekseevDM265} that the problem of finding a maximum
independent set in a graph $G$ of ${\cal G}_3$ is NP-complete,
which completes the proof. \qed

\2

In the rest of this section we will use the notation of Figure
\ref{III_graphs} for the target graph $H.$ We denote by
$\alpha(G)$ the maximal number of vertices in an independent
vertex set of a graph $G.$ We will prove the following lemma using
a  reduction from the problem of finding a maximum independent set
in a $3$-partite graph.

\begin{lemma}\label{bi-claw}
If $H$ is a bipartite claw, then MinHOM($H$) is NP-complete.
\end{lemma}
\pf Let $H$ be a bipartite claw, with
$V(H)=\{x_1,x_2,x_3,x_4,y_1,y_2,y_3\}$ and
$E(H)=\{x_4y_1,y_1x_1,x_4y_2,y_2x_2, x_4y_3, y_3x_3\}$ (see Figure
\ref{III_graphs} (a)). Let $G$ be a $3$-partite graph, with
partite sets $V_1,V_2,V_3$. We will now build a graph $G^*$ for
which mch$(G^*,H)=|V(G)|-\alpha(G)$. This will prove the lemma, by
Theorem \ref{MaxIndepNPhard}.

Let  $G^*$ be obtained from $G$ by inserting a new vertex $m_e$
into every edge $e \in E(G)$. Note that $V(G^*)=V(G) \cup \{ m_e
\mbox{ $|$ } e \in E(G)\}$ and $E(G^*)=\{u m_{uv}, m_{uv} v \mbox{
$|$ } uv \in E(G) \}$.  Define costs as follows, where  $i \in
\{1,2,3\}$ and $j \in \{1,2,3,4\}$.

\2

$\begin{array}{llcll} c_{x_i}(u)=0 & \mbox{if $u \in V_i$} &
\hspace{1.4cm}
  & c_{x_4}(u)=1 & \mbox{ if $u \in V(G)$} \\
c_{x_i}(u)=|V(G)| & \mbox{if $u \not\in V_i$} &
  & c_{y_i}(u)=|V(G)| & \mbox{ if $u \in V(G)$} \\
c_{y_i}(m_e)=0 & \mbox{if $e \in E(G)$} &
  & c_{x_j}(m_e)=|V(G)| & \mbox{if $e \in E(G)$} \\
\end{array} $

\2

Let $I$ be an independent set in $G$, and define a mapping $f$
from $V(G^*)$ to $V(H)$ as follows. For all $u \in V_i$ let
$f(u)=x_i$ if $u \in I$ and $f(u)=x_4$ if $u \not\in I$.  Let $uv
\in E(G)$ be arbitrary, and let $f(m_{uv})=y_i$ if $\{u,v\} \cap
(I \cap V_i)\neq \emptyset$, and let $f(m_{uv})=y_1$ if $x,y
\not\in I$. Note that $f$ is a homomorphism of $G^*$ to $H$ with
cost $|V(G)|-|I|$.

Let $f$ be a homomorphism of $G^*$ to $H$ of cost $|V(G)|-k$. We
will now show that there exists an independent set, $I$ in $G$ of
order at least $k$. If $k\leq 0$ then we are trivially done so
assume that $k>0$, which implies that all individual costs in
$c(f)$ are either zero or one. Let $I=\{u \in V(G) \mbox{ $|$ }
c_{f(u)}(u)=0 \}$ and note that $|I| \geq k$. Note that $I$ is an
independent set in $G$, as if $uv \in E(G)$, where $u \in I \cap
V_i$ and $v \in I \cap V_j$ ($i \not= j$), then $f(u)=x_i$ and
$f(v)=x_j$ which implies that $f$ is not a homomorphism, a
contradiction. Therefore $I$ is independent in $G$.

Observe that we have proved that mch$(G^*,H)=|V(G)|-\alpha(G)$.
Thus, we have now reduced the problem in Theorem
\ref{MaxIndepNPhard} to MinHOM($H$), which completes the proof.
\qed

\2

In the proofs of the next two lemmas, we will again use reductions
from the problem of finding a maximum independent set in a
$3$-partite graph.

\begin{lemma}\label{bi-net}
If $H$ is a bipartite net, then MinHOM($H$) is NP-complete.
\end{lemma}
\pf Let $H$ be a bipartite net, with
$V(H)=\{x_1,x_2,x_3,x_4,y_1,y_2,y_3\}$ and
$E(H)=\{x_1y_1,y_1x_3,y_1x_4,x_3y_2,x_4y_2,y_2x_2,y_3x_4\}$ (see
Figure \ref{III_graphs} (b)). Let $G$ be a $3$-partite graph, with
partite sets $V_1,V_2,V_3$. We will now build a graph $G^*$ such
that mch$(G^*,H)=2|V_3|+|V(G)|-\alpha(G)$. This will prove the
lemma, by Theorem \ref{MaxIndepNPhard}.

Let  $G^*$ be obtained from $G$ in the following way. For every
vertex $v \in V_3$ let $P_v=s_1^vt_1^vs_2^vt_2^vs_3^v$ be a path
of length $4$. For every $u\in V_1$ and $v\in V_2$ with $uv\in
E(G)$ we introduce a new vertex $m_{uv}$. We set $$V(G^*)=V_1 \cup
V_2\cup \{m_e \mbox{ $|$ } e\in E(G)\} \cup \{ V(P_v) \mbox{ $|$ }
v \in V_3 \}.$$ The edge set of $G^*$ consists of the following
edges. For every edge $uv$ between $V_1$ and $V_2$ in $G$ both
$um_{uv}$ and $vm_{uv}$ belong to $G^*$. All edges in $V(P_v)$,
where $v \in V_3$, belong to $G^*$. For all $u \in V_1$ and $v \in
V_3$, where $uv \in E(G)$, the edge $u s_1^{v}$ belongs to $G^*$.
 For all $u \in V_2$ and $v \in V_3$, where
$uv \in E(G)$, the edge $u s_3^{v}$ belongs to $G^*$.

\2

We now define the costs of mapping vertices from $V_1 \cup V_2$ as
follows, where all costs not shown are given the value
$2|V_3|+|V(G)|$. For each $u\in V_i$, $i=1,2$, we set
$c_{x_i}(u)=0$ and $c_{x_4}(u)=1.$ We define the costs of mapping
vertices from $V(G^*)-V_1-V_2$ as follows, where $i \in \{1,2,3\}$
and $j \in \{1,2\}$. For each $e\in E(G)$ and $z\in V(H)$, we set
$c_z(m_e)=0$. Finally, for each $v \in V_3$, we set

\2

$\begin{array}{ccccc}
  c_{y_3}(s_i^v)=0 & \mbox{ and } & c_{q}(s_i^v)=1  & \mbox{ for all } & q \in V(H)-y_3; \\
                        c_{x_4}(t_j^v)=1 & \mbox{ and } & c_{q}(t_j^v)=0 & \mbox{ for all } & q \in
                        V(H)-x_4.
                         \\
\end{array} $

\2

Let $I$ be an independent set in $G$, and define a mapping $f$
from $V(G^*)$ to $V(H)$ as follows. For each $i=1,2$ and $u\in
V_i$,  let $f(u)=x_i$ if $u \in I$ and $f(u)=x_4$ if $u \not\in
I$. For every edge $uv$ of $G$ with $u\in V_1$ and $v\in V_2$, let
$f(m_{uv})=y_2$ if $v\in I$ and $f(m_{uv})=y_1$, otherwise. For
all $v \in V_3 \cap I$ let $f(s_1^v)=f(s_2^v)=f(s_3^v)=y_3$ and
$f(t_1^v)=f(t_2^v)=x_4$. For all $v \in V_3 - I$ let
$f(s_1^v)=f(s_2^v)=y_1$, $f(s_3^v)=y_2$ and
$f(t_1^v)=f(t_2^v)=x_3$. Note that $f$ is a homomorphism of $G^*$
to $H$ with cost $2|V_3|+|V(G)|-|I|$.

Let $f$ be a homomorphism from $G^*$ to $H$ of cost
$2|V_3|+|V(G)|-k$. We will now show that there exists an
independent set $I$ in $G$ of order at least $k$. If $k\leq 0$
then we are trivially done so assume that $k>0$, which implies
that all individual costs in $c(f)$ are either zero or one. Define
$I$ as follows.
$$I=\{u \in V_1 \cup V_2 \mbox{ $|$ } c_{f(u)}(u)=0 \} \cup
    \{v \in V_3 \mbox{ $|$ } f(s_1^v)=f(s_3^v)=y_3 \} $$

We will now show that $I$ is independent in $G$ and that $|I| \geq
k$. First suppose that $uv \in E(G)$, where $u \in I \cap V_i$ and
$v \in I \cap V_j$ ($i \not= j$). Observe that this is not
possible if $\{i,j\}=\{1,2\}$, so without loss of generality
assume that $i<j=3$. However if $i=1$ then we cannot have both
$f(u)=x_1$ and $f(s_1^y)=y_3$ and if $i=2$ then we cannot have
both $f(u)=x_2$ and $f(s_3^y)=y_3$. Therefore $I$ is independent.

If we could show that the cost of mapping $P_v$ to $H$ (denoted by
$c(P_v)$) fulfills (a) and (b) below, then we would be done, as
this would imply that $|I| \geq k$.

\2

$\begin{array}{cclcr}
(a) & &  c(P_v) \geq 2 & \hspace{0.5cm} & \mbox{if $v \in I \cap V_3$} \\
(b) & &  c(P_v) \geq 3 & \hspace{0.5cm} & \mbox{if $v \in V_3-I$} \\
\end{array} $

\2

Indeed,
\begin{eqnarray*}
c(f) & = & \sum_{u\in V_1\cup V_2}c_{f(u)}(u)+\sum_{v\in V_3}c(P_v)\\
     & \ge & (|V_1\cup V_2|-|(V_1\cup V_2)\cap I|) + 2 |V_3\cap I| +3
     (|V_3|-|V_3\cap I|)\\
     & = & 2|V_3|+|V(G)|-|I|
\end{eqnarray*}
and, thus, $|I|\ge k.$

\2

To prove (a) and (b) assume that $v \in V_3$ is arbitrary. Note
that $c_{f(s_1^v)}(s_1^v)>0$ or $c_{f(t_1^v)}(t_1^v)>0$ (or both),
as if $f(s_1^v)=y_3$ then we must have $f(t_1^v)=x_4$. Analogously
$c_{f(s_3^v)}(s_3^v)>0$ or $c_{f(t_2^v)}(t_2^v)>0$ (or both). This
proves (a). If $c_{f(s_2^v)}(s_2^v)>0$, then $c(P_v) \geq 3$, so
assume that $c_{f(s_2^v)}(s_2^v)=0$, which implies that
$f(s_2^v)=y_3$. Thus, $f(t_1^v)=f(t_2^v)=x_4$. If $v \not\in I$
then we have $c_{f(s_1^v)}(s_1^v)>0$ or $c_{f(s_3^v)}(s_3^v)>0$,
which together with $c_{f(t_1^v)}(t_1^v)=c_{f(t_2^v)}(t_2^v)=1$,
implies (b). \qed

\2

\begin{lemma}\label{bi-tent}
If $H$ is a bipartite tent, then MinHOM($H$) is NP-complete.
\end{lemma}
\pf Let $H$ be a bipartite tent with
$V(H)=\{x_1,x_2,x_3,x_4,y_1,y_2,y_3\}$ and
$E(H)=\{x_4y_1,y_1x_1,x_1y_2,y_2x_4, x_1y_3, y_3x_2, x_2y_1,
y_1x_3\}$ (see Figure \ref{III_graphs} (c)). Let $G$ be a
$3$-partite graph, with partite sets $V_1,V_2,V_3$. We will now
build a graph $G^*$ such that mch$(G^*,H)=|V(G)|-\alpha(G)$. This
will prove the lemma, by Theorem \ref{MaxIndepNPhard}.

Let $E_{1,2}$ denote all edges between $V_1$ and $V_2$ in $G$. A
graph $G^*$ is obtained from $G$, by inserting a new vertex
$m_{e}$ into every edge $e \in E_{1,2}$. Note that $V(G^*)=V(G)
\cup \{ m_{e} \mbox{ $|$ } e \in E_{1,2} \}$. The edge set of
$G^*$ consists of all edges in $G$ incident with a vertex in $V_3$
as well as of the edges $ \{u_1 v_{u_1u_2}, v_{u_1u_2} u_2 \mbox{
$|$ } u_1u_2 \in E_{1,2} \}$. We now define the costs of $u_i \in
V_i$ as follows, where all costs not shown are given the value
$|V(G)|$.

\2

$\begin{array}{lclcl}
\mbox{For $i=1$:} & \hspace{0.1cm} & c_{y_2}(u_1)=0 & \hspace{0.4cm} & c_{y_1}(u_1)=1 \\
\mbox{For $i=2$:} & \hspace{0.1cm} & c_{y_3}(u_2)=0 & \hspace{0.4cm} & c_{y_1}(u_2)=1 \\
\mbox{For $i=3$:} & \hspace{0.1cm} & c_{x_3}(u_3)=0 & \hspace{0.4cm} & c_{x_1}(u_3)=1 \\
\end{array} $

\2

For all edges $e \in E_{1,2}$ let $c_{x_1}(m_e)=|V(G)|$ and let
$c_{q}(m_e)=0$ for all $q \in V(H)-\{x_1\}$.

\2

  Let $I$ be an independent set in $G$, and define a mapping $f$ from
$V(G^*)$ to $V(H)$ as follows.

\2

\begin{tabular}{llcll}
 For $u \in V_1 \cap I$: & $f(u)=y_2$ & \hspace{1.2cm} & For $u \in V_1 - I$: & $f(u)=y_1$ \\
 For $u \in V_2 \cap I$: & $f(u)=y_3$ &                & For $u \in V_2 - I$: & $f(u)=y_1$ \\
 For $u \in V_3 \cap I$: & $f(u)=x_3$ &                & For $u \in V_3 - I$: & $f(u)=x_1$ \\
\end{tabular}

\2

If $u_1u_2 \in E_{1,2}$ and $u_1 \in V_1 \cap I$, then let
$f(m_{u_1u_2})=x_4$. If $u_2 \in V_2 \cap I$, then let
$f(m_{u_1u_2})=x_2$. If $u_1,u_2 \not\in I$ then let
$f(m_{u_1u_2})=x_4$. Note that $f$ is a homomorphism from $G^*$ to
$H$ with cost $|V(G)|-|I|$.

Let $f$ be a homomorphism from $G^*$ to $H$ of cost $|V(G)|-k$. We
will now show that there exists an independent set, $I$ in $G$ of
order at least $k$. If $k\leq 0$ then we are trivially done so
assume that $k>0$, which implies that all individual costs in $f$
are either zero or one. Let $I=\{u \in V(G) \mbox{ $|$ }
c_{f(u)}(u)=0 \}$ and note that $|I| \geq k$. Furthermore, observe
that $I$ is an independent set in $G$ (as $f(v_e) \not=x_1$ for
every $e \in E_{1,2}$). We have reduced the problem in Theorem
\ref{MaxIndepNPhard} to MinHOM$(H)$, which completes the proof.
\qed

\vspace{2mm}

\begin{corollary}
If $H$ is a connected irreflexive graph which is not a proper
interval bigraph, then MinHOM$(H)$ is NP-complete.
\end{corollary}

\pf If $H$ is not bipartite, this follows from the fact that the
homomorphism problem for $H$ is NP-complete \cite{hellJCT48}.
Otherwise, the conclusion now follows from Theorem \ref{mainnp}.
\qed

\vspace{2mm}

Since we have observed that a connected $H$ which contains both
loops and nonloops gives rise to an NP-complete problem
MinHOM$(H)$, it only remains to prove the NP-completeness of
MinHOM$(H)$ when $H$ is a reflexive graph which is not a proper
interval graph. There is an analogous result characterizing proper
interval graphs by the absence of induced cycles of length at
least four, or a claw, net, or tent
\cite{wegner,golumbic2004,spin}. However, we instead reduce the
problem to the irreflexive case, as follows.

Given a reflexive graph $H$, we define the bipartite graph $H^*$
with the vertex set $\{v', v'':\ v \in V(H)\}$ and edge set
$\{v'v'':\ v \in V(H)\} \cup \{u'v'':\ uv \in E(H)\}$. It is
proved in \cite{hellJGT} that $H$ is a proper interval graph if
and only if $H^*$ is a proper interval bigraph. Thus suppose a
reflexive graph $H$ is not a proper interval graph, and consider
the bipartite (irreflexive) graph $H^*$ which is then not a proper
interval bigraph. We will now reduce the NP-complete problem
MinHOM$(H^*)$ to the problem MinHOM$(H)$ as follows. Each instance
of MinHOM$(H^*)$ can also be viewed as an instance of MinHOM$(H)$.
Indeed, such an instance consists of a bipartite graph $G$ with
costs $c_{i'}(u)$ for each white vertex $u$ of $G$ and white
vertex $i'$ of $H^*$, and costs $c_{i''}(v)$ for each black vertex
$v$ of $G$ and black vertex $i''$ of $H^*$; to see this as an
instance of MinHOM$(H)$, we only need to set $c_i(u)$ equal to
$c_{i'}(u)$ if $u$ is white and $c_{i''}(u)$ if $u$ is black. Now
colour-preserving homomorphisms of $G$ to $H^*$ and to $H$ are in
a one-to-one correspondence, with the same costs, i.e., there is a
homomorphism of $G$ to $H^*$ of cost not exceeding $k$ if and only
if there is a homomorphism of $G$ to $H$ of cost not exceeding
$k$.

\begin{corollary}
If a connected graph $H$ with loops allowed is neither a reflexive
proper interval graph nor an irreflexive proper interval bigraph,
then the problem MinHOM$(H)$ is NP-complete.
\end{corollary}

\section{Digraphs}\label{frsec}

A digraph $H$ (with loops allowed) satisfying the Min-Max property
yields a polynomial time solvable problem MinHOM$(H)$ (Theorem
\ref{MMd}). However, there are other digraphs $H$ for which the
problem MinHOM$(H)$ admits a polynomial solution. For instance, it
is easy to see that when $H$ is a directed cycle, we can solve
MinHOM$(H)$ in polynomial time, cf. \cite{gutinDAM}. On the other
hand, the directed cycle $\vec{C}_p$ clearly does not have the
Min-Max property, as can be seen by considering the vertex $w_p$
and its two incident edges.

The classification problem for the complexity of minimum cost
digraph homomorphism problems remains open. However, in
\cite{gutinDO}, a partial classification has been obtained for the
class of {\em semicomplete $k$-partite digraphs}. These are
digraphs that can be obtained from undirected complete $k$-partite
graphs by orienting each undirected edge in one direction or in
both directions. When $k \geq 3$, the classification in
\cite{gutinDO} is completed. When $k=2$, it is only completed when
no edge is oriented in both directions. The authors of
\cite{gutinDO} have remarked there that a certain of these
problems are polynomially equivalent to minimum cost homomorphism
problems to undirected bipartite graphs. Those problems have been
classified here, settling one additional family of digraph
homomorphism problems to semicomplete bipartite digraphs. However,
the full classification of this case is still open, as is the
general family of all minimum cost digraph homomorphism problems.
On the other hand, dichotomy of {\em list} homomorphism problems
for digraphs follows from a recent result of Bulatov
\cite{bulatovACMTCL}.

\2

\2

{\bf Acknowledgements} Research of Gutin and Rafiey was supported
in part by the IST Programme of the European Community, under the
PASCAL Network of Excellence, IST-2002-506778.

{\small

\end{document}